\documentclass[11pt]{article}
\normalbaselines
\usepackage{psfig}

\newcounter{eqnletter}[equation]
\catcode`\@=11
\@addtoreset{equation}{section}   
\catcode`\@=12

\begin{document}

\pagestyle{empty}

\begin{center}

{\LARGE\bf RECTANGULAR WELL AS PERTURBATION\\[2cm]}

\vskip 2cm

{\large {\bf Mariusz Dudek , 
	Stefan Giller{$\dag$}\footnote{Supported by KBN 2PO3B 07610}}}\\
{\bf and}\\ 
{\large {\bf Piotr Milczarski{$\ddag$}\footnote
{Supported by the {\L}\'od\'z University Grant No 580}}}

\vskip 1cm
Theoretical Physics Department II, University of {\L}\'od\'z,\\
Pomorska 149/153, 90-236 {\L}\'od\'z, Poland \\
e-mail: $\dag$ sgiller@krysia.uni.lodz.pl \\
	$\ddag$ jezykmil@krysia.uni.lodz.pl

\end{center}
\vspace{1 cm}
\begin{abstract}

We discuss a finite rectangular well as a perturbation
for the infinite one with a depth $\lambda^2$ of the former as a
perturbation parameter. In particular we consider a behaviour of
energy levels in the well as functions of complex $\lambda$. It is found
that all the levels of the same parity are defined on infinitely
sheeted Riemann surfaces which topological structures are
described in details.  These structures differ considerably from
those found in models investigated earlier \cite{1,4,8,9,10}. It is
shown that perturbation series for all the levels converge what
is in a contrast with the known results of Bender and Wu \cite{1}.
The last property is shown to hold also for the infinite
rectangular well with Dirac delta barrier as a perturbation
considered earlier by Ushveridze \cite{4}.  

\end{abstract}

\vspace{1cm}

\begin{tabular}{l}
{\small PACS number(s): 03.65.-W } \\[5mm]
\end{tabular}

\newpage

\pagestyle{plain}

\setcounter{page}{1}

\section{Introduction}

Since the papers of Bender and Wu \cite{1} we have known
why the perturbation series were in general divergent.  We have known
also that in many cases investigations of perturbation series
could be reduced to the investigations of the corresponding
semiclassical series \cite{2}. It was also realized that the
divergent perturbation series could be summed and one of the
summation methods applied here was very often the Borel one
\cite{2,3}.

One of byproducts of these investigations was a
discovery of so called level crossing i.e. of the fact that in
the case of confining polynomial potentials all the discrete
energy levels they produce or only groups of them are no longer
isolated of each other if considered as functions of a
perturbation parameter \cite{1,3,8,9} i.e. the levels inside each
group appear as branches of a ramified functions of the
perturbation parameter considered as a complex variable. Being
more precise the latter statement means that each energy level
belonging to a group considered as a function of a real
perturbation parameter can be analytically continued into the
complex plane of the parameter so that any energy level of the
group can be reached by the analytic continuation procedure of
some arbitrary chosen level belonging to the group. This means
also that the complex plane of the perturbation parameter
converts rather into some (more or less) complicated Riemann
surface.  

It is also well known \cite{7} that it is an existing
symmetry group of the Hamiltonian considered which is completely
responsible for a decay of the energy spectrum into disjoint
(with respect to analytic continuation) groups of them.
Therefore, a degree of complication of respective Riemann
surfaces on which the energy levels are defined can give us an
information about an existence of the relevant symmetry group
i.e. the more levels appear as branches defined on respective
sheets of the same Riemann surface the less rich a relevant
symmetry group has to be. In particular if the corresponding
Riemann surfaces are all finitely sheeted (i.e. if there are
finite numbers of energy levels attached to each of them) then they have
to be defined by some algebraic conditions relating energies and
a perturbation parameter, with the conditions being a clear sign
of the existence of an underlying symmetry group \cite{7}.  

There are only a few examples of the analysis described above in the case
when the relevant Riemann surface is infinitely sheeted \cite{1,4}.
The analysis performed is more or less numerical. This is mostly
because it is very difficult to find an example of potential
providing us with a closed functional form of the quantization
condition being simultaneously sufficiently simple to perform an
analysis in a 'classical', non numerical way. According to these
investigations a Riemann surface topology corresponding to a
given perturbation parameter i.e. loci of its branch points
seems to be still mysterious and depending on a parameter chosen
so needing still further studies.  

In this paper we will consider a possibly simple but non trivial example of an
Hamiltonian provided by the familiar rectangular well of a
{\it finite} height which allows us for such a classical analysis. As
a perturbation parameter in this example we will choose its
height $V=\lambda^2>0$; strictly speaking the square root of it). The
perturbed potential is then the infinite rectangular well
approached when $V^{-1}\rightarrow 0$. 
There are several basic properties which differ
the case of the {\it finite} rectangular well from the ones considered
earlier. First it is just a finite number of energy levels
existing for a given $\lambda$ but varying with $\lambda$ so that a 
potentially
infinite number of levels can appear when $\lambda^{-1}\rightarrow 0$.
The remaining properties are enumerated as points 2., 4. and 5. below.  

The rectangular well is not an analytical potential and as such it
provides us also with a non analytical quantization condition.
However an analytical extension of the latter into complex
values of the quantities considered is possible and results of
the relevant analysis are the following: 

1. The system of energy levels of the well decays into two disjoint families 
(of different parities) with the levels inside each of the group
being analytical continuations of each other with respect to the
perturbation parameter; 

2. The perturbation series for each level is {\bf convergent} to the level itself 
i.e. the property which is quite opposite to that of Bender and Wu for the 
unharmonic oscillator case. As such they are trivially Borel summable; 

3. The Riemann surfaces for both the groups of levels are
infinitely sheeted and their branch point structures can be
understood by some simple properties of both the quantization
conditions.  

4. The energy level poles existing in the complex
momentum plane corresponding to the case are accompanied by
poles which do not represent nor discrete nor resonant parts of
the energy spectrum (in particular because resonances are absent
in the case of the finite rectangular well independently of
whether the latter is a real well (for real $\lambda$) or is a rectangular
barrier (for imaginary $\lambda$)). These second sort of poles we shall
call pseudoenergy levels.  

5. The level crossing which happens for {\it real} $\lambda $ is not between 
two real  energies but just between an energy and its pseudoenergy partner.  

We will reconsider also an example of a numerical analysis performed earlier 
by Ushveridze \cite{4} to show that its 'classical' analysis is possible in 
its full size confirming the main results of the author mentioned
but completed them with such an important conclusion as a
convergence of both perturbation series corresponding to the
weak and the strong couplings to the Dirac delta perturbation
used in the example.  

Another goal of our investigations was
looking for rules governing distributions of energy level branch
points on the perturbation parameter Riemann surface. It has
been demonstrated by Ushveridze \cite{4} that one can predict an
existence of energy level crossing as well as arrange the
crossing to join more than two levels but as in the case of the
anharmonic potential \cite{1} distributions of the corresponding
branch points seemed to be unpredictable a priori to direct
calculations.  

Our main conclusion in this respect is similar.
Indeed such distributions of branch points is strongly model
(potential) dependent and it seems that only some crude rules
can be formulated to predict their presence or absence in some
domains of the underlying Riemann surface.

\section{Finite rectangular well as a perturbation} 

\vspace{5mm}

{\it {\bf 2.1 Analytic properties of energy levels as functions of 
perturbation parameter }}

\vspace{5mm}

A finite rectangular well as a perturbation sounds a
little bit exotically but it can be considered as such for the
infinite one in the same way as for example a finite potential well: 
\begin{eqnarray}
\frac{1}{2}
\frac{\omega^2}{\alpha^2}
\left( 1-\frac{1}{\cosh({\alpha}x)} \right),
\label{1}
\end{eqnarray}
being a perturbation for the harmonic one i.e. we get the latter from
(\ref{1}) for $\alpha \rightarrow 0$.  

In the case of a finite rectangular potential given by: 
\begin{eqnarray}
V(x) =
\left\{
\begin{array}{ll}
V(>0) & \mbox{for $\mid x \mid >1$} \\[3mm]
0 & \mbox{for $\mid x \mid <1$}
\end{array}
\right.
\label{2}
\end{eqnarray}
a perturbation parameter can be chosen to be $\lambda^2=V$ which together
with the rescaled energy $E/V \equiv z^2$ leads us to the following
quantization conditions \cite{5}: 
\begin{eqnarray}
\left\{
\begin{array}{llll}
\lambda \mid \cos{\phi} \mid = \phi, & z = \mid \cos{\phi} \mid , 
& \tan{\phi}>0 & \mbox{for positive parity levels} \\[3mm]
\lambda \mid \sin{\phi} \mid = \phi, & z = \mid \sin{\phi} \mid , 
& \tan{\phi}<0 & \mbox{for negative parity levels}
\end{array}
\right.
\label{3}
\end{eqnarray}

All (positive) solutions $\phi_{k}(\lambda)$ ,
$k=$0,1,2,..., to Eq.'s (\ref{3}) are represented picturesquely in Fig.1
as given by the points of intersections of the straight line 
${\phi}/{\lambda}$ with the right arcs of $\mid sin(\phi) \mid$ and 
$\mid \cos(\phi) \mid$ functions. The corresponding
solutions for energy levels are then get as 
$E_{k}(\lambda)={\lambda^2}{z^2}({\phi}_{k}(\lambda))$.
It is just a dependence of $E_{k}$'s on $\lambda$ considered as 
a complex parameter which is the main interest of this paper.  

\begin{tabular}{c}
\psfig{figure=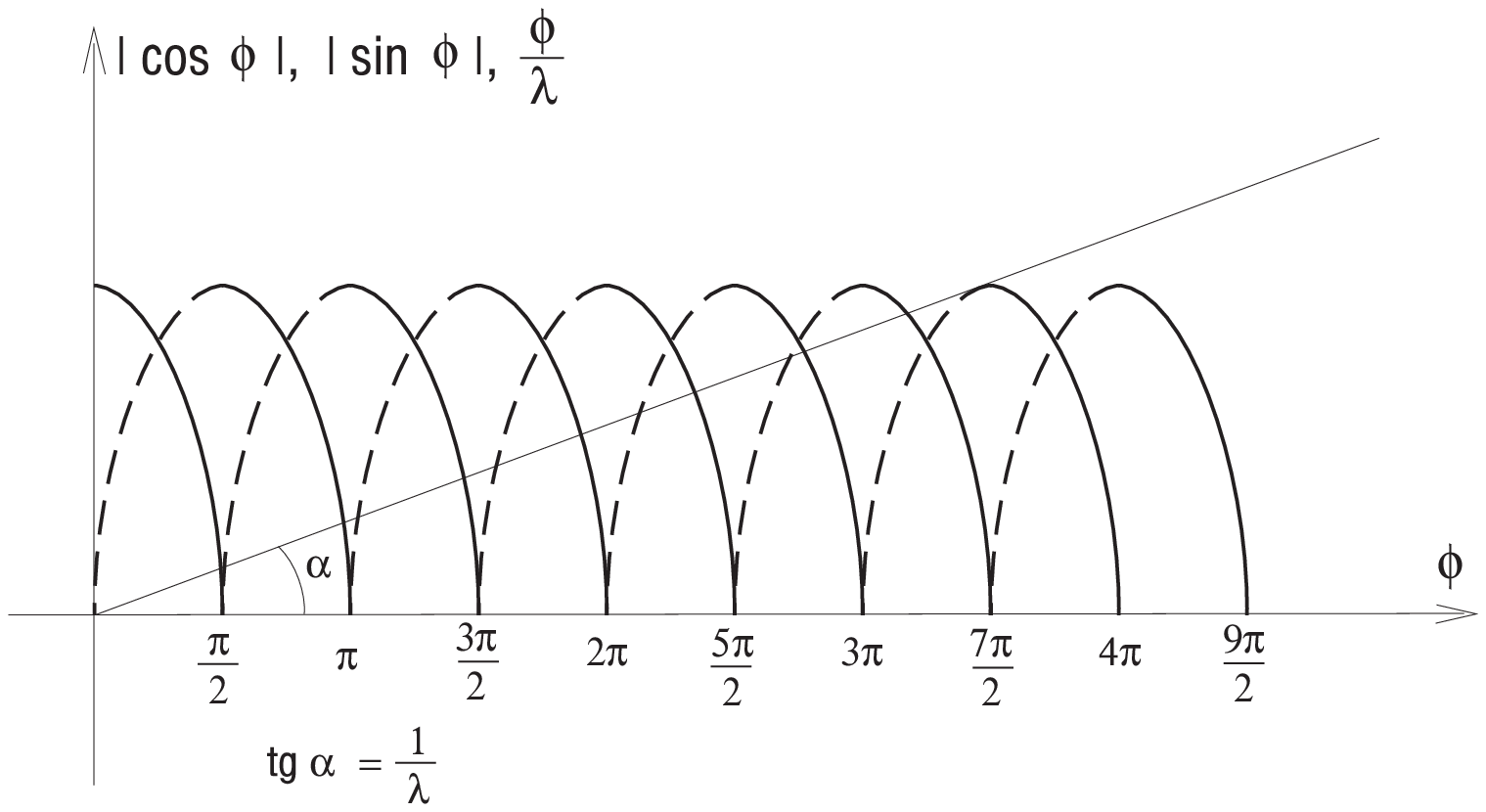,width=10cm} \\
\parbox{14cm}{Fig.1 The $\phi$ angles chosen by the quantization
conditions (2.3)}
\end{tabular}

\vspace{5mm}

To investigate this dependence it is however necessary to make an analytical 
continuation of the conditions (\ref{3}) into the complex $\lambda$. 
Such a continuation of Eq.'s (\ref{3}) is possible when dropping the
absolute value marks in (\ref{3})
what provides us with following analytic conditions: 

\begin{eqnarray}
\left\{
\begin{array}{llll}
\lambda \cos{\phi} = \phi ,& z = \cos{\phi} , 
& \mbox{for $0 \leq \phi \leq \frac{1}{2}\pi$} ,& \mbox{(mod $2\pi$)} \\[3mm]
\lambda \cos{\phi} = - \phi ,& z = - \cos{\phi} ,
& \mbox{for $\pi \leq \phi \leq \frac{3}{2} \pi$} ,& \mbox{(mod $2\pi$)} \\[3mm]
\lambda \sin{\phi} = \phi ,& z = \sin{\phi} , 
& \mbox{for $\frac{1}{2}\pi \leq \phi \leq \pi$} ,& \mbox{(mod $2\pi$)} \\[3mm]
\lambda \sin{\phi} = - \phi ,& z = - \sin{\phi} , 
& \mbox{for $\frac{3}{2} \pi \leq \phi \leq 2\pi$} ,& \mbox{(mod $2\pi$)}
\end{array}
\right.
\label{4}
\end{eqnarray}
equivalent to (\ref{3}) for positive $\phi$.  

According to (\ref{4}) the energy spectrum for
the potential (\ref{2}) is formally divided into four groups. Each
group contains every forth members of the spectrum starting from
$E_{1}^{+}$ , $E_{3}^{+}$ , $E_{2}^{-}$ , $E_{4}^{-} $
energy levels in the corresponding group. The energy levels in
the groups are defined by the conditions (\ref{4}) in the order
mentioned.  

Dropping further the restrictions for the ranges of
changing $\phi$ in the conditions (\ref{4}) we get fully analytical
quantization conditions but describing rather different spectra
with respect to which our original ones are only parts of them.
This is however the necessary price for investigating the
complex analytical dependence of energy levels on $\lambda$.  

When doing it however we observe that for both the parity levels it is
enough to continue analytically only the firsts of the
corresponding conditions (\ref{4}) depriving them of the corresponding
restrictions for $\phi$ (i.e. $\phi$ can now take any complex value in these
conditions).  This is because corresponding solutions $\phi^{\pm}(\lambda)$ 
to the first conditions generate solutions $\phi^{\pm}(-\lambda)$ to the second ones. 
(In fact for the even parity both the solutions almost coincide since
$\phi^{+}(-\lambda)=-\phi^{+}(\lambda)$).  
Therefore we see that all energy levels of both parities: 
$E_{k}^{+}(\lambda)=\lambda^{2}cos^2{\phi_{k}^{+}}(\lambda)$
and
$E_{k}^{-}(\lambda)=\lambda^{2}sin^2{\phi_{k}^{-}}(\lambda)$
can be obtained by solving only the first conditions (\ref{4}) of the
corresponding parities and performing analytic continuations in $\lambda$
from its positive to its negative values.  

We will analyze first the even parity group determining the ground state 
energy level. An analysis of the odd parity case is quite similar.  

\vspace{5mm}
{\it {\bf a) Even parity energy spectrum case }}

\vspace{5mm}
Making in the first of the
conditions (\ref{4}) a change of variable $e^{i\phi} = \sigma$ we get instead:
\begin{eqnarray}
\left\{
\begin{array}{l}
\lambda^{+}= - \frac{2i{\sigma}ln{\sigma}}{\sigma^{2}+1} \\[5mm]
z^{+} = \frac{1}{2} \left( \sigma + \frac{1}{\sigma} \right)
\end{array}
\right.
\label{5}
\end{eqnarray}

Since a dependence of $z^{+}$  on $\sigma$ as given by (\ref{4}) is rather 
simple then 
to get the corresponding dependence of $z^{+}$ on $\lambda$ it is necessary 
to invert the dependence of the latter variable on $\sigma$ as shown in 
(\ref{5}). To do this one needs to know: 

$1^{0}$ The Riemann surface structure for $\lambda^{+}(\sigma)$ ;

$2^{0}$ The loci of all zeros of $\lambda^{+'}(\sigma)$ on the surface; and 

$3^{0}$ A pattern of lines $Re{\lambda^{+}}=const$ 
and $Im{\lambda^{+}}=const$
on the surface.  

As it follows from (\ref{5}) the Riemann
surface structure for ${\lambda}^{+}(\sigma)$ is determined by: 

a. The logarithmic branch point at $\sigma = 0$ ; and 

b. A pair of simple poles at $\sigma = {\pm} i$ located on
every sheet the latter being generated in an infinite number by
the logarithm.  

Zeros of $\lambda^{+'}(\sigma)$ are determined by the following
equation:
\begin{eqnarray}
ln{\sigma} = \frac{{\sigma}+1}{{\sigma}-1}
\label{6}
\end{eqnarray}

Putting $e^{i \phi + y} = \sigma$ ($\sigma$ is now an arbitrary complex 
number on the surface) and assuming that $\phi,y\neq0$ we transform (\ref{6}) 
into:
\begin{eqnarray}
\left\{
\begin{array}{l}
\frac{sin2{\phi}}{2\phi} = - \frac{sinh2y}{2y} \\[5mm]
cos{2\phi} = - \frac{sinh2y}{2y} + cosh2y \\[5mm]
\phi , y \neq 0
\end{array}
\right.
\label{7}
\end{eqnarray}
from which it follows that all zeros of ${\lambda}^{+'}(\sigma)$ 
have to lie on the circle $\mid{\sigma}\mid = 1$ 
and/or on the real half axis $\phi = 0$.  
Therefore the corresponding conditions for them are:
\begin{eqnarray}
\left\{
\begin{array}{l}
cot{\phi} = - \phi \\[5mm]
ln{\sigma} = \frac{{\sigma}+1}{{\sigma}-1}
\end{array}
\right.
\label{8}
\end{eqnarray}

Solutions to the first of Eq.'s (\ref{8}) (in fact infinitely many of
them) are given therefore as the intersection points of the
functions cot$\phi$ and $-\phi$. The form of the condition can be easily
identified as the one for the straight line ${\phi}/{\lambda}$ of Fig. 1 to be
tangent to cos$\phi$. Therefore the points have to lie close to $\phi=2k\pi$ 
ones, $k=$1,2,..., on the left to them, and close to $\phi=k\pi$, $k=$-1,-3,..., 
on the right to the latter.  

Two real solutions to the second of
Eq.'s (\ref{8}) are placed on both the sides of $\sigma=1$ at 
$\sigma_{1} \approx 3.32$ and at $\sigma_{1}^{-1} \approx 0.301$.  

Let us note that the points on the unit circle 
at $\phi=k\pi , k=-1,+2,-3,+4,....$,
and the point $\sigma=1$ are all the physical
'thresholds' for successive appearing of the corresponding even
parity energy levels according to changing $\lambda$ from zero to
infinity. There is a temptation to understand the zeros provided
by the conditions (\ref{8}) as a shifting of these thresholds from
their real physical positions mentioned to their actual ones
because of the approximations which the analytical conditions
(\ref{8}) effectively are to our rectangular well quantization
problem. This shifting of thresholds remains in a deep relation
to analytical properties of reflection and transmission
coefficients of the corresponding scattering problem arising
when energy is higher than $\lambda^2$. We will discuss this relation
below.  

Although the solutions to the condition (\ref{8}) cannot be
considered as thresholds for any real energy spectrum case (this
is excluded by the absence of the energy level degeneracy in
1-dim SE) we shall consider them as such for convenience and
call them pseudothresholds.  

Thus the solutions to the first of
the conditions (\ref{8}) are the pseudothresholds for the energy
levels lying inside the potential above the ground state one.
These pseudothresholds as we have mentioned earlier coincide
with results of demanding for the line ${\phi}/{\lambda}$ in Fig. 1 
to be tangent to $cos{\phi}$. The latter demand is just the lower limit for 
$\lambda$ above
which the real solutions for the energy levels can exist. Both
the solutions at $\sigma_{1}$ and $\sigma_{1}^{-1}$  between which  
$\lambda^{+}(\sigma)$ is pure imaginary
are then the limits for the ground state energy level $E_{1}^{+}$ to be
a real quantity. However for $E_{1}^{+}(\lambda)$ to be real is not enough to
represent a real physical level.  The sufficient conditions will
be discussed below in section 2.2.  

A pattern of the lines $Re{\lambda}^{+}=Im{\lambda}^{+}=0$ is sketched 
on Fig.2 together with the positions of
all singular points of $\lambda^{+}(\sigma)$. Since all zeros of 
$\lambda^{+'}(\sigma)$ are simple
they result as square root branch point singularities of $\sigma(\lambda)$ on
its $\lambda$-Riemann surface. A consequence of this is a perpendicular
crossing of two lines $Im{\lambda}^{+}=0$ as well as the corresponding two
lines $Re{\lambda}^{+}=0$ in every of such zero. All the lines 
$Im{\lambda}^{+}=0$
emanating outside the unit circle run to infinity on the
corresponding sheets of the $\sigma$-Riemann surface parallel
asymptotically to the line $Re{\sigma}=0$.  These emanating inside the
circle have to cross the logarithmic branch point at $\sigma=0$ being
tangent at this point to the line $Re{\sigma}=0$. The remaining lines
$Im{\lambda}^{+}=0$ coincide with all semicircles of the unit circle lying
between the poles at $\sigma= \pm i$ on every logarithmic sheet of the
$\sigma$-Riemann surface.  

\begin{tabular}{c}
\psfig{figure=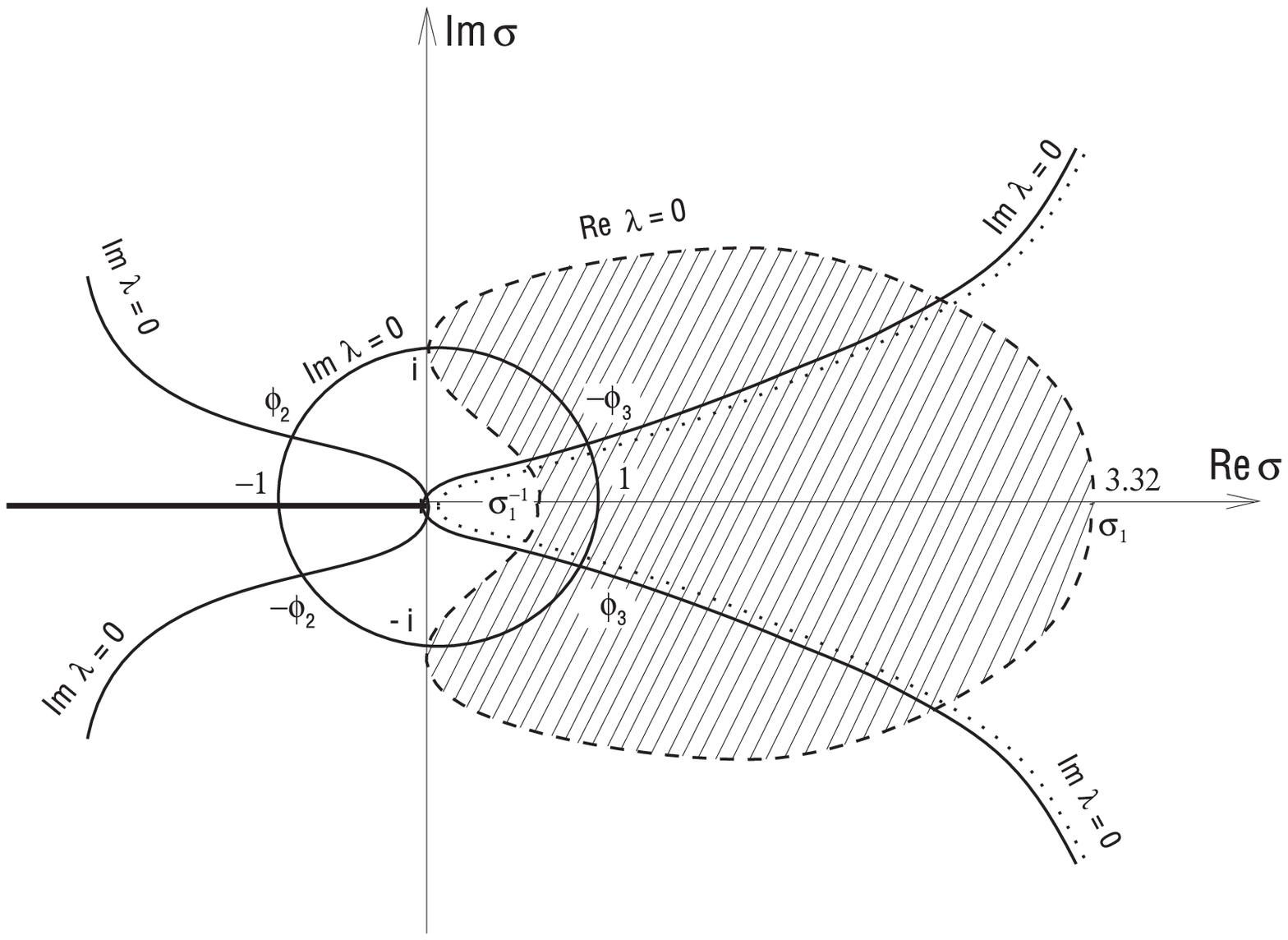,width=10cm} \\
\parbox{14cm}{Fig.2 The 'first' sheet of the $\sigma$-Riemann
surface corresponding to $\lambda^{+}(\sigma)$} 
\end{tabular}

\vspace{5mm}

The $\lambda$-Riemann surface on which the inverse
function $\sigma^{+}(\lambda)$ is defined is now easy to construct. 
It is also an
infinitely sheeted surface. Its first sheet is a map of the
dashed area in Fig. 2 and is shown in Fig. 3a. This is the sheet
on which the ground state energy level is defined i.e. for $\lambda$
changing along the real axis of the sheet or along a segment
$[{\lambda}_{1},-{\lambda}_{1}]{\equiv}[-{\lambda}^{+}({\sigma}_{1}),
{\lambda}^{+}({\sigma}_{1}^{-1})]$
on the imaginary axis where the corresponding
energy ${E}_{1}^{+}$ is real. On the rest of the sheet the energy is
complex.  

There are two cuts on the sheet emanating of the two
complex conjugate imaginary branch points at 
${\lambda}_{1}={\lambda}^{+}({\sigma}_{1})=-{\lambda}^{+}({\sigma}_{1}^{-1})$
and running to infinity along its imaginary axis. The function 
${\sigma}^{+}(\lambda)$
is holomorphic on the sheet approaching ${\pm}i$ for $\lambda$ escaping to the
infinity on the right or on the left half planes of the sheet
respectively.  The second sheet corresponds to the second energy
level. The sheet can be achieved by crossing (in any direction)
one of the two cuts described above. Despite these two latter
cuts there are another two square root branch points on the
sheet lying on the real axis at 
${\lambda}={\pm}{\lambda}_{2}={\pm}{{\lambda}^{+}}(e^{\textstyle{-i{\phi_{2}}}})={\mp}{\phi}_{2}/cos{\phi}_{2}$. 
The energy
level ${E}_{2}^{+}$ is given by the values of $\lambda$ changing on the both 
sides of the real axis from the real branch points mentioned up to the
corresponding infinities. On the $\sigma$-Riemann surface these ranges
of changing $\lambda$ correspond to varying $\sigma$ along the two left unit
semicircles between the poles at $\sigma={\pm}i$.

There is still an
additional branch point on the discussed sheet at $\lambda=0$ which is
also present at all the other sheets except the first one. It is
a common picture of the point $\sigma=0$ and all the infinity points of
the $\sigma$-Riemann surface of Fig. 2. Therefore this point gives rise
to an infinite branching of $\lambda^{+}(\sigma)$ around it.

Starting from the
third energy level every of the levels is represented on two
different sheets which can communicate with themselves by the
branch point at $\lambda=0$. Each of these two sheets belongs to two
different groups of them the latter being generated by the two
branch points of the second sheet at ${\lambda}={\pm}{\lambda}_{2}$ 
(see Fig. 3b). Let us
analyze the group generated by the branch point at ${\lambda}={\lambda}_{2}$. 
The other
group is a twin picture of this one obtained by a transformation 
${\lambda}{\rightarrow}-{\lambda}$. 

Going around ${\lambda}_{2}$ through the cut $A_{2}/{\bar{B_{2}}}$ we find 
ourselves on the sheet shown in Fig. 3c. There is an additional cut on the
sheet generated by the branch point at 
${\lambda}={\lambda}_{3}={{\lambda}^{+}}(e^{\textstyle{-i{\phi_{3}}}})=-{\phi}_{3}/cos{\phi}_{3}$.
The branch point opens possibility for the fourth and the fifth
energy levels to appear. The ranges of the energy levels $E_{3}^{+}$ and
$E_{4}^{+}$ are shown on the figure.  

\begin{center}
\begin{eqnarray*}
\unitlength=0.60mm
\special{em:linewidth 0.4pt}
\linethickness{0.4pt}
\begin{picture}(121.00,125.00)
\put(75.00,30.00){\line(0,1){90.00}}
\put(75.00,123.00){\makebox(0,0)[ct]{$\wedge$}}
\put(30.00,75.00){\line(1,0){90.00}}
\put(121.00,75.00){\makebox(0,0)[cc]{$>$}}
\put(87.00,125.00){\makebox(0,0)[rt]{Im$\lambda$}}
\put(118.00,82.00){\makebox(0,0)[ct]{Re$\lambda$}}
\put(88.00,83.00){\makebox(0,0)[ct]{$E_{1}^{+}$}}
\put(63.00,83.00){\makebox(0,0)[ct]{$E_{1}^{+}$}}
\put(83.00,97.00){\makebox(0,0)[rb]{$\lambda_{1}$}}
\put(86.00,53.00){\makebox(0,0)[rc]{$-\lambda_{1}$}}
\put(69.00,110.00){\makebox(0,0)[cc]{$A_{1}$}}
\put(81.00,110.00){\makebox(0,0)[cc]{$\bar{B_{1}}$}}
\put(81.00,40.00){\makebox(0,0)[cc]{$\bar{B_{1}^{'}}$}}
\put(69.00,40.00){\makebox(0,0)[cc]{$A_{1}^{'}$}}
\put(33.00,118.00){\makebox(0,0)[cc]{a)}}
\put(74.00,97.00){\rule{2.00\unitlength}{23.00\unitlength}}
\put(74.00,30.00){\rule{2.00\unitlength}{23.00\unitlength}}
\end{picture}
&
\unitlength=0.60mm
\special{em:linewidth 0.4pt}
\linethickness{0.4pt}
\begin{picture}(121.00,125.00)
\put(75.00,30.00){\line(0,1){90.00}}
\put(75.00,123.00){\makebox(0,0)[ct]{$\wedge$}}
\put(30.00,75.00){\line(1,0){90.00}}
\put(121.00,75.00){\makebox(0,0)[cc]{$>$}}
\put(87.00,125.00){\makebox(0,0)[rt]{Im$\lambda$}}
\put(118.00,82.00){\makebox(0,0)[ct]{Re$\lambda$}}
\put(83.00,97.00){\makebox(0,0)[rb]{$\lambda_{1}$}}
\put(86.00,53.00){\makebox(0,0)[rc]{$-\lambda_{1}$}}
\put(69.00,110.00){\makebox(0,0)[cc]{$B_{1}$}}
\put(81.00,110.00){\makebox(0,0)[cc]{$\bar{A_{1}}$}}
\put(81.00,40.00){\makebox(0,0)[cc]{$\bar{A_{1}^{'}}$}}
\put(69.00,40.00){\makebox(0,0)[cc]{$B_{1}^{'}$}}
\put(60.00,74.00){\rule{30.00\unitlength}{2.00\unitlength}}
\put(74.00,95.00){\rule{2.00\unitlength}{25.00\unitlength}}
\put(74.00,30.00){\rule{2.00\unitlength}{25.00\unitlength}}
\put(81.00,81.00){\makebox(0,0)[cc]{$A_{2}$}}
\put(69.00,81.00){\makebox(0,0)[cc]{$A_{2}^{'}$}}
\put(81.00,69.00){\makebox(0,0)[cc]{$\bar{B_{2}}$}}
\put(69.00,69.00){\makebox(0,0)[cc]{$\bar{B_{2}^{'}}$}}
\put(56.00,69.00){\makebox(0,0)[cc]{$-\lambda_{2}$}}
\put(94.00,69.00){\makebox(0,0)[cc]{$\lambda_{2}$}}
\put(100.00,81.00){\makebox(0,0)[cc]{$E_{2}^{+}$}}
\put(50.00,81.00){\makebox(0,0)[cc]{$E_{2}^{+}$}}
\put(35.00,120.00){\makebox(0,0)[cc]{b)}}
\end{picture}
\\
\unitlength=0.60mm
\special{em:linewidth 0.4pt}
\linethickness{0.4pt}
\begin{picture}(121.00,125.00)
\put(75.00,30.00){\line(0,1){90.00}}
\put(75.00,123.00){\makebox(0,0)[ct]{$\wedge$}}
\put(30.00,75.00){\line(1,0){90.00}}
\put(121.00,75.00){\makebox(0,0)[cc]{$>$}}
\put(87.00,125.00){\makebox(0,0)[rt]{Im$\lambda$}}
\put(118.00,82.00){\makebox(0,0)[ct]{Re$\lambda$}}
\put(60.00,74.00){\rule{30.00\unitlength}{2.00\unitlength}}
\put(81.00,81.00){\makebox(0,0)[cc]{$B_{2}$}}
\put(69.00,81.00){\makebox(0,0)[cc]{$A_{3}$}}
\put(81.00,69.00){\makebox(0,0)[cc]{$\bar{A_{2}}$}}
\put(69.00,69.00){\makebox(0,0)[cc]{$\bar{B_{3}}$}}
\put(56.00,69.00){\makebox(0,0)[cc]{$-\lambda_{3}$}}
\put(94.00,69.00){\makebox(0,0)[cc]{$\lambda_{2}$}}
\put(100.00,81.00){\makebox(0,0)[cc]{$E_{3}^{+}$}}
\put(50.00,81.00){\makebox(0,0)[cc]{$E_{4}^{+}$}}
\put(35.00,120.00){\makebox(0,0)[cc]{c)}}
\end{picture}
&
\unitlength=0.60mm
\special{em:linewidth 0.4pt}
\linethickness{0.4pt}
\begin{picture}(121.00,125.00)
\put(75.00,30.00){\line(0,1){90.00}}
\put(75.00,123.00){\makebox(0,0)[ct]{$\wedge$}}
\put(30.00,75.00){\line(1,0){90.00}}
\put(121.00,75.00){\makebox(0,0)[cc]{$>$}}
\put(87.00,125.00){\makebox(0,0)[rt]{Im$\lambda$}}
\put(118.00,82.00){\makebox(0,0)[ct]{Re$\lambda$}}
\put(60.00,74.00){\rule{30.00\unitlength}{2.00\unitlength}}
\put(81.00,81.00){\makebox(0,0)[cc]{$A_{4}$}}
\put(69.00,81.00){\makebox(0,0)[cc]{$B_{3}$}}
\put(81.00,69.00){\makebox(0,0)[cc]{$\bar{B_{4}}$}}
\put(69.00,69.00){\makebox(0,0)[cc]{$\bar{A_{3}}$}}
\put(56.00,69.00){\makebox(0,0)[cc]{$-\lambda_{3}$}}
\put(94.00,69.00){\makebox(0,0)[cc]{$\lambda_{4}$}}
\put(100.00,81.00){\makebox(0,0)[cc]{$E_{6}^{+}$}}
\put(50.00,81.00){\makebox(0,0)[cc]{$E_{5}^{+}$}}
\put(35.00,120.00){\makebox(0,0)[cc]{d)}}
\end{picture}
\end{eqnarray*} \\

Fig.3 The few first $\lambda$-Riemann surface sheets for $\sigma^{+}(\lambda)$
 $(E^{+}(\lambda))$
\end{center}

\vspace{5mm}

To achieve a sheet corresponding
to a pair $E_{5}^{+}$ , $E_{6}^{+}$  of the levels we have to cross the cut 
$A_{3}/{\bar{B_{3}}}$
in Fig. 3c (in any direction). The sheet is shown in Fig. 3d
together with corresponding ranges for the energies.  

A full structure of the considered group of sheets of the $\lambda$-Riemann
surface is now obvious. The $n^{th}$ sheet, n=3,4,..., contains three
branch points lying at $\lambda=0$ , 
${\lambda}={\lambda}_{n-1}=-{\phi}_{n-1}/cos{\phi}_{n-1}$
 and at ${\lambda}={\lambda}_{n}=-{\phi}_{n}/cos{\phi}_{n}$.
Each of the last two branch points opens a pair of energy
levels: $E_{2n-3}^{+}$, $E_{2n-2}^{+}$ and $E_{2n-1}^{+}$ , $E_{2n}^{+}$ , 
respectively, the levels in each pair lying on different sheets. 
And inversely, the energy levels appear on every sheet in pairs. 
A $2n^{th}$ sheet corresponding to energy levels 
$E_{4n-1}^{+}$ , $E_{4n}^{+}$ , $n=1,2$,..., (see
Fig. 4) is cut by two cuts: 
${B_{2n}}/{\bar{A}_{2n}}$ beginning at 
${\lambda}={\lambda}_{2n}=-{\phi}_{2n}/cos{{\phi}_{2n}}>0$ 
and opening the level $E_{4n-1}^{+}$ and ${B_{2n+1}}/{\bar{A}_{2n+1}}$
beginning at 
${\lambda}={\lambda}_{2n+1}=-{\phi}_{2n+1}/cos{{\phi}_{2n+1}}<0$
 and opening the level $E_{4n}^{+}$.
Both the cuts end at $\lambda=0$. A corresponding sheet for the levels
$E_{4n+1}^{+}$, $E_{4n+2}^{+}$ is cut from $\lambda_{2n+1}$ to 0 and from 0 
to ${\lambda}={\lambda}_{2n+2}=-{\phi}_{2n+2}/cos{{\phi}_{2n+2}}>0$
with the latter branch point opening the
second of the considered levels (see Fig. 4).  

\begin{center}
\begin{eqnarray*}
\unitlength=0.76mm
\special{em:linewidth 0.4pt}
\linethickness{0.4pt}
\begin{picture}(92.00,96.00)
\put(46.00,1.00){\line(0,1){90.00}}
\put(46.00,94.00){\makebox(0,0)[ct]{$\wedge$}}
\put(1.00,46.00){\line(1,0){90.00}}
\put(92.00,46.00){\makebox(0,0)[cc]{$>$}}
\put(58.00,96.00){\makebox(0,0)[rt]{Im$\lambda$}}
\put(89.00,53.00){\makebox(0,0)[ct]{Re$\lambda$}}
\put(31.00,45.00){\rule{30.00\unitlength}{2.00\unitlength}}
\put(52.00,52.00){\makebox(0,0)[cc]{$B_{2n}$}}
\put(40.00,52.00){\makebox(0,0)[cc]{$A_{2n+1}$}}
\put(52.00,40.00){\makebox(0,0)[cc]{$\bar{A}_{2n}$}}
\put(40.00,40.00){\makebox(0,0)[cc]{$\bar{B}_{2n+1}$}}
\put(27.00,40.00){\makebox(0,0)[cc]{$\lambda_{2n+1}$}}
\put(65.00,40.00){\makebox(0,0)[cc]{$\lambda_{2n}$}}
\put(71.00,52.00){\makebox(0,0)[cc]{$E_{4n-1}^{+}$}}
\put(21.00,52.00){\makebox(0,0)[cc]{$E_{4n}^{+}$}}
\put(81.00,81.00){\makebox(0,0)[cc]{$2n^{th}$}}
\end{picture}
&
\unitlength=0.76mm
\special{em:linewidth 0.4pt}
\linethickness{0.4pt}
\begin{picture}(92.00,96.00)
\put(46.00,1.00){\line(0,1){90.00}}
\put(46.00,94.00){\makebox(0,0)[ct]{$\wedge$}}
\put(1.00,46.00){\line(1,0){90.00}}
\put(92.00,46.00){\makebox(0,0)[cc]{$>$}}
\put(58.00,96.00){\makebox(0,0)[rt]{Im$\lambda$}}
\put(89.00,53.00){\makebox(0,0)[ct]{Re$\lambda$}}
\put(31.00,45.00){\rule{30.00\unitlength}{2.00\unitlength}}
\put(52.00,52.00){\makebox(0,0)[cc]{$A_{2n+2}$}}
\put(40.00,52.00){\makebox(0,0)[cc]{$B_{2n+1}$}}
\put(52.00,40.00){\makebox(0,0)[cc]{$\bar{B}_{2n+2}$}}
\put(40.00,40.00){\makebox(0,0)[cc]{$\bar{A}_{2n+1}$}}
\put(27.00,40.00){\makebox(0,0)[cc]{$\lambda_{2n+1}$}}
\put(65.00,40.00){\makebox(0,0)[cc]{$\lambda_{2n+2}$}}
\put(71.00,52.00){\makebox(0,0)[cc]{$E_{4n+2}^{+}$}}
\put(21.00,52.00){\makebox(0,0)[cc]{$E_{4n+1}^{+}$}}
\put(81.00,81.00){\makebox(0,0)[cc]{$2n+1^{th}$}}
\put(46.00,46.00){\oval(70.00,24.00)[]}
\put(62.00,58.00){\makebox(0,0)[cc]{$<$}}
\put(32.00,34.00){\makebox(0,0)[cc]{$>$}}
\put(72.00,63.00){\makebox(0,0)[cc]{C}}
\end{picture}
\end{eqnarray*} \\
Fig.4 The general structure of $\lambda$-Riemann surface sheets for 
$\sigma^{+}(\lambda)$ $(E^{+}(\lambda))$
\end{center}

The structure of the second group of sheets generated by the branch 
point at ${\lambda}={-\lambda_{2}}$ 
is obtained by an inversion: $\lambda$ $\rightarrow$ $-\lambda$ from 
the first one discussed above.

\vspace{5mm}

{\it {\bf b) Odd parity energy spectrum case }}
\vspace{5mm}

Making in the third of the
conditions (\ref{4}) a change of variable ${\sigma}=e^{i{\phi}}$ 
we get the condition in the form :

\begin{eqnarray}
\left\{
\begin{array}{l}
\lambda^{-} =  \frac{2{\sigma}ln{\sigma}}{\sigma^{2}-1} \\[5mm]
z^{-} = \frac{1}{2i} (\sigma - \frac{1}{\sigma})
\end{array}
\right.
\label{9}
\end{eqnarray}

It follows from (\ref{9}) that ${\lambda}^{-}(\sigma)$ is a meromorphic 
function of $\sigma$ on the
$\sigma$-Riemann surface with the logarithmic branch point at $\sigma=0$ and
with simple poles at $\sigma={\pm}1$ on every logarithmic sheet except the
first one where the pole at $\sigma=1$ is absent. The surface is shown
in Fig. 5 where the lines $Im{\lambda^{-}}(\sigma)=0$ are also 
shown schematically.
Zeros of ${\lambda^{-'}}(\sigma)$ are distributed in this case only on 
the unit
circle $\mid \sigma \mid =1$ at the point $\sigma = 1$ on the 
first sheet and at points $\sigma = e^{\textstyle i{\phi_{k}}}$
with $\phi_{k}(=-{\phi_{-k}}>0)$ satisfying the condition 
$\phi_{\pm k} = \tan{\phi_{\pm k}}$ , $k=2,...$ , so that  the first 
two of the latter singular points lie on the second sheet of Fig. 5.  

\begin{tabular}{c}
\psfig{figure=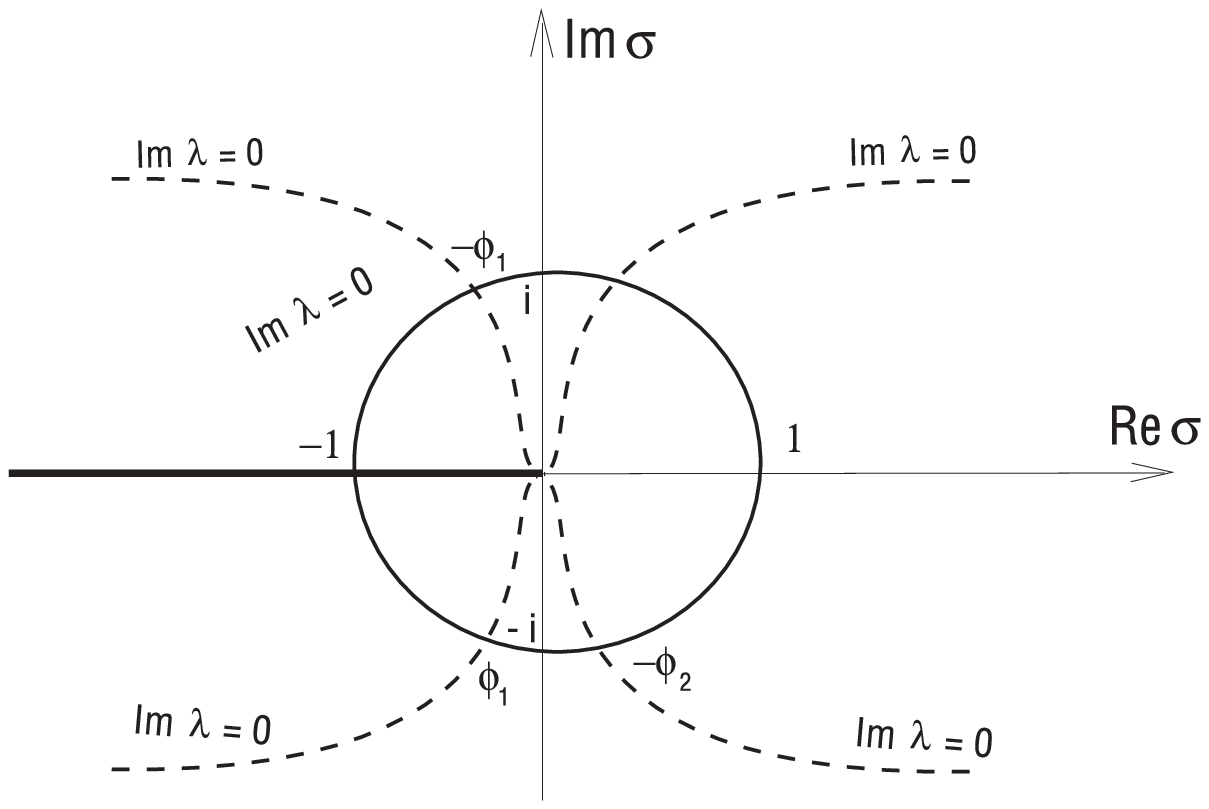,width=9cm} \\
\parbox{14cm}{Fig.5 The 'first' sheet of the $\sigma$-Riemann
surface corresponding to $\lambda^{-}(\sigma)$ }
\end{tabular}

\vspace{5mm}

On the other hand the inverse function
$\sigma^{-}(\lambda)$ is a holomorphic one on the 
$\lambda$-Riemann surface with branch
points at $\lambda = 1$ and at $\lambda_k = \lambda^{-}(e^{i{\phi_k}})$ 
which are images of zeros of $\lambda^{-'}(\sigma)$.

The spectrum is opened with the level $E_{2}^{-}(\lambda)$ 
which belongs to the
odd parity spectrum of the rectangular well opening it contrary
to the next level $E_{3}^{-}(\lambda)$ which does not.

An interesting property of 
$E_{2}^{-}(\lambda)$ is a singularity it has to have at $\lambda = 0$ 
i.e. the level $E_{2}^{-}(\lambda)$ 
as a function of$ \lambda$ is not bounded from below at this point.
This conclusion follows from an observation that although the
point $\lambda = 1$ is a branch one for $\sigma^{-}(\lambda)$ 
(below which $\sigma^{-}(\lambda)$ has two
values: $\sigma$ and $\sigma^{-1}$ for every $\lambda$ , 
$0 \leq \lambda \leq 1$) it is not as such for the level
$E_{2}^{-}(\lambda)$. On the other hand $E_{2}^{-}(\lambda)$ 
$(=-[\ln{\sigma}(\sigma^{2} + 1)/(\sigma^{2} + 1)]^{2})$ 
for $0<{\lambda}<1$ ($0<{\sigma}<+{\propto}$)
is real and negative. Therefore a pseudothreshold for the
latter appears to be at the point $\lambda = 0$ at which 
$E_{2}^{-}(\lambda)$ becomes
infinitely large and negative.  Here again we want to stress
however that the realness of $E_{2}^{-}(\lambda)$ does not mean automatically
its existence as a real physical energy level (see a discussion
below, Sec.  2.3) and in the case considered $E_{2}^{-}(\lambda)$ disappears as
a physical level below $\lambda = {\pi}/2$ as it follows from Fig. 1.  

The remaining levels have already expected properties only.  

The level $E_{3}^{-}(\lambda)$ starts with a pseudothreshold 
$E_{3}^{-}(e^{\textstyle i{\phi_{1}}})$. A sheet on
which both the levels vary is shown in Fig. 6. It emerges as a
map of an area of Fig. 5. lying between the real halfline $\sigma {\geq} 0$ 
and the line $Im{\lambda^{-}}(\sigma)=0$ crossing the point 
$\sigma = e^{\textstyle i{\phi_{1}}}$. 
The map is provided by $\lambda^{-}(\sigma)$.  

The second sheet and all the next ones are arranged in a
very similar way to the corresponding sheets in the even parity
energy level case. To achieve the second sheet on which the
levels $E_{4}^{-}(\lambda)$ and $E_{5}^{-}(\lambda)$ are defined 
we should cross a cut $A_{2}/{\bar{B_{2}}}$
of Fig. 6 in any direction close to the branch point
$\lambda_{1}={\lambda^{-}}(e^{\textstyle i{\phi_{1}}}) = {\phi_{1}}/{\sin{\phi_{1}}}<0$. 
The point opens the level $E_{4}^{-}(\lambda)$ on the sheet
for $-\propto < {\lambda} < {\lambda_{1}}$. 
There is a second branch point on the sheet at
$\lambda = \lambda_{2} = {\phi_{2}}/{\sin{\phi_{2}}}>0$.  
opening the level $E_{5}^{-}(\lambda)$ for $\lambda_{2} < \lambda < +\propto$. 
The sheet emerges
as a map of an area of Fig. 5 lying between two successive lines
$Im{\lambda^{-}}(\sigma)=0$, the first one crossing the singular point 
$\sigma_{1} = e^{\textstyle i{\phi_{1}}}$ and the
next one crossing the unit circle at 
$\sigma_{2} = e^{\textstyle i{\phi_{2}}}$.  

All the next sheets
appear as maps of successive areas bounded by corresponding
pairs of the lines $Im{\lambda^{-}}(\sigma)=0$ crossing the 
successive singular points 
$\sigma_{k} = e^{\textstyle i{\phi_{k}}}$, $k=2$,3,..., 
in the direction of an increasing
argument of $\sigma$. However, contrary to the even parity case maps of
the negative argument sheets of the $\sigma$-Riemann surface do not
produce additional (twin in forms) sheets on the $\lambda$-Riemann
surface what corresponds to the absence of relevant right cut on
the first sheet of the latter surface. It means that the complex
conjugated positive and negative argument sheets of the $\sigma$-Riemann
surface map into the same sheets of the $\lambda$-Riemann ones for the
odd energy function $E^{-}(\lambda)$.

\begin{tabular}{c} 
\unitlength=0.72mm
\special{em:linewidth 0.4pt}
\linethickness{0.4pt}
\begin{picture}(96.00,99.00)
\put(50.00,4.00){\line(0,1){90.00}}
\put(50.00,97.00){\makebox(0,0)[ct]{$\wedge$}}
\put(5.00,49.00){\line(1,0){90.00}}
\put(96.00,49.00){\makebox(0,0)[cc]{$>$}}
\put(62.00,99.00){\makebox(0,0)[rt]{Im$\lambda$}}
\put(93.00,56.00){\makebox(0,0)[ct]{Re$\lambda$}}
\put(35.00,48.00){\rule{30.00\unitlength}{2.00\unitlength}}
\put(44.00,55.00){\makebox(0,0)[cc]{$A_{2}$}}
\put(44.00,43.00){\makebox(0,0)[cc]{$\bar{B_{2}}$}}
\put(31.00,43.00){\makebox(0,0)[cc]{$\lambda_{1}$}}
\put(69.00,43.00){\makebox(0,0)[cc]{$1$}}
\put(75.00,55.00){\makebox(0,0)[cc]{$E_{2}^{-}$}}
\put(25.00,55.00){\makebox(0,0)[cc]{$E_{3}^{-}$}}
\end{picture}
\\
\parbox{14cm}{Fig.6 The sheet of the $\lambda$-Riemann surface
corresponding to the level 
$E_{2}^{-}(\lambda)$ and its pseudolevelcompartner $E_{3}^{-}(\lambda)$ }
\end{tabular}

\vspace{5mm}

\vspace{5mm}

{\it {\bf 2.2 Relation between analytical properties of energy levels and
analytical properties of transmission coefficient }}

\vspace{5mm}

It is a standard result of the 1-dim quantum mechanics that energy
levels of bound states are simple poles for a transmission
coefficient of an accompanied 1-dim scattering problem \cite{5}.
These poles have to occupy the positions on the positive
imaginary axis of the complex momentum corresponding to an
infinite motion. In fact this last property is the main
criterion for selecting the real bound state energies from the
whole set of poles the transmission coefficient can have.  Since
the analysis of the analytical quantization conditions of the
previous section provided us with a variety of the solutions
which all have to be poles for a transmission coefficient then
to select out of them those being the physical bound state
levels we have to discuss roles played by them in the
corresponding transmission coefficient.  

We will give to our
considerations a standard formulation shifting all energy levels
by $-\lambda^2$ so that the infinite motion takes place for $E>0$ with the
momentum $k={\sqrt{E}}$ outside the well and with the momentum 
$k'=(k^{2} + {\lambda^2})^{1/2}$
inside it (the bottom of the well is now at $V = - \lambda^2$, of course).
Then the reflection (R) and transmission (T) coefficients for
the case are the following:

\begin{eqnarray}
T(k) =  \frac{kk'e^{-2ik}}{(k{\cos{k'}}-ik'{\sin{k'}})({k'}{\cos{k'}}-ik{\sin{k'}})} 
\label{10}
\end{eqnarray}
\begin{eqnarray*}
R(k) = \frac{i{\lambda^2}{\sin{2k'}}}{2kk'} T(k)
\end{eqnarray*}

It is seen from (\ref{10}) that T (and R as well) as a function of
complex {\it k} is meromorphic with their poles coinciding with roots
of the T-denominator. Clearly these poles occupy exactly the
positions of energy levels in the {\it k}-plane we have found in the
previous paragraphs. The denominator factorization in (\ref{10})
occurs due to the reflection symmetry of the potential well. In
particular the first denominator factor in (\ref{10}) corresponds to
the odd parity levels and the second to the even ones.  

We can separate the investigations of the levels of different parities
(according to that we have done earlier) by considering instead
of the coefficients (\ref{10}) two pairs of the following ones:
\begin{eqnarray}
\left\{
\begin{array}{l}
T^{+}(k) = \frac{k}{{k'}{\cos{k'}}-ik{\sin{k'}}} \\[5mm]
R^{+}(k) = \frac{i{\lambda}{\cos{k'}}}{k} T^{+}(k) \\[5mm]
T^{-}(k) = \frac{k}{k{\cos{k'}}-i{k'}{\sin{k'}}} \\[5mm]
R^{-}(k) = \frac{i \lambda \sin{k'}}{k} T^{-}(k)
\end{array}
\right.
\label{11}
\end{eqnarray}

Of course, the two above pairs of coefficients correspond now to
two different potential wells: $V^{+}(x,{\lambda})$ and 
$V^{-}(x,{\lambda})$ respectively
which can be reconstructed with the help of the inverse
scattering method \cite{6}.

Let us consider first a pole structure
of the {\it k}-plane corresponding to the even parity case. The
corresponding analysis of the odd one is analogous.  

This structure is shown in Fig. 7 and follows from an observation
that $\phi_{l}^{+}(\lambda)$ as a $l^{th}$ analytical solution 
to the first of the
conditions (\ref{4}) defines a pole of $T^{+}(k)$ on the {\it k}-plane at
$k_{l}^{+}(\lambda) = i \lambda \sin{\phi_{l}^{+}(\lambda)}$.

\begin{tabular}{c}
\unitlength=0.9mm
\special{em:linewidth 0.4pt}
\linethickness{0.4pt}
\begin{picture}(112.00,122.00)
\put(55.00,122.00){\makebox(0,0)[cc]{$\wedge$}}
\put(112.00,80.00){\makebox(0,0)[cc]{$>$}}
\put(110.00,73.00){\makebox(0,0)[cb]{Rek}}
\put(61.91,119.92){\makebox(0,0)[cc]{Imk}}
\put(46.94,110.11){\makebox(0,0)[cc]{$i\lambda$}}
\put(46.94,49.93){\makebox(0,0)[cc]{$-i\lambda$}}
\put(54.93,49.93){\makebox(0,0)[cc]{$\bullet$}}
\put(54.93,110.11){\makebox(0,0)[cc]{$\bullet$}}
\put(54.93,56.04){\makebox(0,0)[cc]{$\bullet$}}
\put(54.93,104.00){\makebox(0,0)[cc]{$\bullet$}}
\put(54.93,71.95){\makebox(0,0)[cc]{$\bullet$}}
\put(54.93,85.03){\makebox(0,0)[cc]{$\bullet$}}
\put(54.93,64.10){\makebox(0,0)[cc]{$\vdots$}}
\put(54.93,91.06){\makebox(0,0)[cc]{$\vdots$}}
\put(42.94,71.08){\makebox(0,0)[cc]{$\bullet{k_{2l-2}^{+}}$}}
\put(73.03,71.08){\makebox(0,0)[cc]{$\bullet{k_{2l-2}^{+}}$}}
\put(78.03,64.08){\makebox(0,0)[cc]{$\bullet{k_{2l-2}^{+}}$}}
\put(38.03,64.08){\makebox(0,0)[cc]{$\bullet{k_{2l-2}^{+}}$}}
\put(85.00,56.00){\makebox(0,0)[cc]{$\ddots$}}
\put(88.00,52.00){\makebox(0,0)[cc]{$\bullet$}}
\put(95.00,48.00){\makebox(0,0)[cc]{$\ddots$}}
\put(98.00,44.00){\makebox(0,0)[cc]{$\bullet$}}
\put(61.00,97.00){\makebox(0,0)[cc]{$k_{3}^{+}$}}
\put(61.00,104.00){\makebox(0,0)[cc]{$k_{1}^{+}$}}
\put(61.00,85.00){\makebox(0,0)[cc]{$k_{2m-1}^{+}$}}
\put(61.00,72.00){\makebox(0,0)[cc]{$k_{2m-2}^{+}$}}
\put(61.00,56.00){\makebox(0,0)[cc]{$k_{2}^{+}$}}
\put(54.93,97.00){\makebox(0,0)[cc]{$\bullet$}}
\put(4.00,80.00){\line(1,0){106.00}}
\put(55.00,24.00){\line(0,1){96.00}}
\put(28.00,57.00){\makebox(0,0)[cc]{$\cdot$}}
\put(27.00,56.00){\makebox(0,0)[cc]{$\cdot$}}
\put(26.00,55.00){\makebox(0,0)[cc]{$\cdot$}}
\put(24.00,52.00){\makebox(0,0)[cc]{$\bullet$}}
\put(15.00,42.00){\makebox(0,0)[cc]{$\bullet$}}
\put(20.00,48.00){\makebox(0,0)[cc]{$\cdot$}}
\put(19.00,47.00){\makebox(0,0)[cc]{$\cdot$}}
\put(18.00,46.00){\makebox(0,0)[cc]{$\cdot$}}
\end{picture} 
\\
\parbox{14cm}{Fig.7 The k-plane of the transmission coefficient
$T^{+}(k)$  }
\end{tabular}

\vspace{5mm}

According to our earlier observation together
with solutions we have for a given $\lambda>0$ we have to consider also
the ones we have for $-\lambda$.

There is a finite number {\it m} 
($\mid{\phi}_{m}^{+}(\pm{\lambda})\mid \leq{m{\pi}}\leq{\lambda}$) 
of solutions to (\ref{4}) for which 
${\phi}_{l}^{+}(\pm{\lambda})$ are real and 
$\pm{\lambda}\sin{\phi}_{m}^{+}(\pm{\lambda})>0$. 
Poles corresponding to these solutions (which include the one for the
ground state (for $l=1$)) are distributed on the segment $(0,i\lambda)$ of
the positive imaginary axis and represent the physical bound
states of the potential $V^{+}(x,{\lambda})$.  

There are also {\it m-1} real solutions to (\ref{4}) but with 
$\pm{\lambda}\sin{\phi}_{m}^{+}(\pm{\lambda})<0$
generating in that way
poles lying in the $k$-plane on the segment ($0,-i$) (i.e. {\it below} the
real axis of the plane).  Therefore the latter which we call
pseudoenergy levels cannot represent the bound states in 
$V^{+}(x,{\lambda})$.
These poles appear as branch partners of the previous ones
(except the ground state partner which does not exist) which
therefore can coincide with the latter at the corresponding
pseudothresholds $\lambda = \lambda_{l}$ (see figures 3-4). 
However, loci of these
pseudothresholds in the $k$-plane are just below the real axis
not contradicting therefore the physical level nondegeneracy
theorem. According to our previous notation (see Fig. 3-4) these
$2m-1$ poles are: 
$k_{1}^{+}(\lambda)$, $(k_{2}^{+}(\lambda),k_{3}^{+}(\lambda))$, 
$(k_{4}^{+}(\lambda),k_{5}^{+}(\lambda))$,...,
$(k_{2m-2}^{+}(\lambda),k_{2m-1}^{+}(\lambda))$
where the branch partners are paired and the
poles with odd indices represent the physical levels (lie above
the real axis).

Finally, there are infinitely many poles of
$T^{+}(k,{\lambda})$ lying below the real axis of the $k$-plane on both the
sides of the imaginary one and symmetrically with respect to it
(due to the relation 
$k_{l}(\lambda)=k_{l}^{*}(\lambda^{*})$
considered for real $\lambda$) but
outside of it with finite distances between any two of them and
because of this distributed up to infinity. These are of course
the poles: 
$(k_{2l-2}^{+}(\lambda),k_{2l-1}^{+}(\lambda))$, $l=m+1,m+2$,...,.  

If $\lambda$ varies
through real positive values increasing all the poles lying
outside the segment $(-i\lambda,i\lambda)$ move towards it 
and a symmetric pair of them 
$(k_{2l-2}^{+}(\lambda),k_{2l-1}^{+}(\lambda))$
achieves the segment just for $\lambda$
approaching $\lambda_{l}$ a value being a corresponding pseudothreshold for
the pair.  While achieving the segment the pair disjoints again
with its member $k_{2l-1}^{+}(\lambda)$
moving upwards imaginary axis and with
the member $k_{2l-2}^{+}(\lambda)$
moving downwards. The first one crosses
eventually the real axis becoming a bound state whilst the
second one does it never becoming a pseudoenergy level.

The above description works for every pair of the pole partners
except the one corresponding to the ground state energy which is
single. This for real positive $\lambda$ (as well as for the negative one
due to a relation $k_{l}(\lambda)=k_{l}(-\lambda))$
valid for any $\lambda$) is above the real
{\it k}-axis and thus represents the ground state energy. However, for
imaginary $\lambda(\sigma)$, when $0 < \sigma < 1$ or 
$1 < \sigma < + \propto$, $E_{1}^{+}(\lambda) = \lambda^2$ 
is real (and of course
negative) but $k_{1}^{+}(\lambda)$ has negative imaginary part and therefore
$E_{1}^{+}(\lambda)$ cannot represent a bound state. This fits well our
intuition since for imaginary $\lambda$ the rectangular well become rather
a rectangular barrier excluding of course any bound state.

A similar analysis of the odd parity levels corresponding to the
potential $V^{-}(x,\lambda)$ leads us to the following conclusions.

For a given $\lambda > 0$ and $m \pi \leq \lambda$ poles: 
$k_{2}^{-}(\lambda)$, 
$(k_{3}^{-}(-\lambda), k_{4}^{-}(-\lambda))$,  
$(k_{5}^{-}(\lambda), k_{6}^{-}(\lambda))$, 
$(k_{7}^{-}(-\lambda),\\ 
k_{8}^{-}(-\lambda))$,...,  
$(k_{2m-1}^{-}((-1)^{m-1}\lambda), k_{2m}^{-}((-1)^{m-1}\lambda))$ 
lie in the 
segment $(-i\lambda,i\lambda)$ whilst the remaining ones
$(k_{2l-1}^{-}((-1)^{l-1}\lambda), k_{2l}^{-}((-1)^{l-1}\lambda))$, 
$l=m+1, m+2$,..., lie outside the 
segment, below the real axis and symmetrically with respect to
imaginary axis.

A behaviour of all the paired poles with
varying along the positive real axis is exactly the same as in
the even parity level case. Only the pole $k_{2}^{-}(\lambda)$ 
seems to differ with this respect in comparison with 
$k_{1}^{-}(\lambda)$ escaping to infinity
along the negative imaginary axis when 
$\lambda \rightarrow 0_{+}$. However if one
considers a behaviour of the levels rather as functions of $\sigma$ 
for $\sigma \rightarrow 0_{+}$
(or for $\sigma \rightarrow +\propto$) 
on the corresponding first sheets of figures 5 and
2 respectively (this corresponds to 
$\lambda \rightarrow 0_{+}$ for the odd level but
describes a little bit more complex path for the even one ending
however at $\lambda=0$) then one can find that both the levels behave
identically.

\vspace{5mm}

{\it {\bf 2.3 Perturbation series for energy levels and their summability }}

\vspace{5mm}

Since the finite rectangular well considered is a perturbation
for the infinite rectangular one when $\lambda^{-2} \rightarrow 0$ 
then the corresponding energy levels of the former should approach the 
corresponding levels of the latter potential in this limit. It is interesting
that with the previous considerations we are able to conclude
that all the corresponding perturbation series expansions are
convergent. The corresponding conclusions are in fact obvious
for all the energy levels higher than the ground state one. To
show this let us consider a pair of levels 
$E_{4n+1}^{+}$, $E_{4n+2}^{+}$ lying
on the $2n+1^{th}$ sheet shown in Fig. 4 with the level $E_{4n+1}^{+}$
belonging to the even spectrum of the finite rectangular well. A
function $E^{+}(\lambda)={\lambda^{2}}z^{2}(\lambda)$ 
being a holomorphic extension of both the
levels considered is holomorphic on the $2n+1^{th}$ sheet outside any
closed contour C containing the branch points 
$\lambda=0$, ${\lambda}_{2n+1}$ and ${\lambda}_{2n+2}$
inside (see Fig. 4). For $\lambda \rightarrow \propto$ on the sheet 
$E^{+}(\lambda) \sim E_{4n+1}^{+}(\lambda) \sim {\pi}^{2}(8n+1)^{2}/4$
and therefore according to the Cauchy theorem we get for $E^{+}(\lambda)$:
\begin{eqnarray}
E^{+}(\lambda) = \frac{((8n+1)\pi)^2}{4} - \frac{1}{2{\pi}i}{{\oint_{C}}
\frac{E^{+}(\lambda')}{{\lambda}'-\lambda}}d{\lambda}'
\label{12}
\end{eqnarray}

It follows from (\ref{12}) that the series
\begin{eqnarray}
\frac{((8n+1)\pi)^2}{4} + \sum_{k{\geq}0} \frac{a_{2n+1,k}}{\lambda^{k+1}} 
\label{13}
\end{eqnarray}
with 
\begin{eqnarray}
a_{2n+1,k} = \frac{(-1)^{k}}{2{\pi}i}{\oint_{C}}E^{+}(\lambda)
{\lambda^{k}}d{\lambda}
\label{14}
\end{eqnarray}
which is the perturbative one for the $E_{4n+1}^{+}(\lambda)$ level is
convergent to the level for 
$\mid \lambda \mid > \mid \lambda_{2n+2} \mid$
i.e. for all $\lambda$'s just above its
pseudothreshold.  

For the ground state energy level $E_{1}^{+}(\lambda)$ an
analogous conclusion can be obtained by considering instead the
cut pattern shown in Fig. 3a the one when the upper and the
lower left cut boundaries in the figure are rotated by $\pm \pi$
respectively providing us with a sheet arranged according to
Fig. 8. It follows from the figure that the perturbative series
for $E_{1}^{+}(\lambda)$ is obtained from (\ref{12}) by putting there $n=1$ 
and $C=C'$ with its convergence radius given by 
$\max(\mid \lambda_{1} \mid , \mid \lambda_{2} \mid)$.  

The same conclusions can be drawn for the perturbative series constructed
for the odd parity energy levels applying exactly the same
technique of considerations as we did in the case of the even
parity levels. The series constructed for a level $E_{n}^{-}(\lambda)$ is
therefore convergent for $\mid \lambda \mid$ sufficiently large on the sheet on
which the level is defined.

\begin{tabular}{c}
\psfig{figure=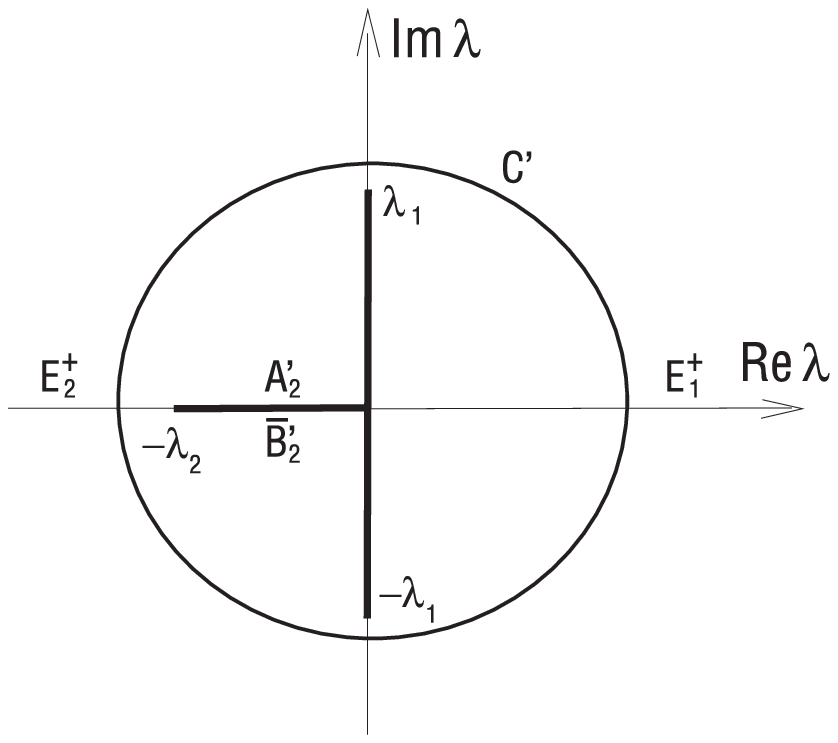,width=9cm} \\
\parbox{14cm}
{Fig.8 The cut structure of the sheet on which the perturbative series
for $E_{1}^{+}(\lambda)$  is studied }
\end{tabular}

\vspace{5mm}

\section{Dirac delta barrier as perturbation}

A second example of a perturbation which can be analyzed
analytically is provided by the Dirac delta barrier introduced
into the infinite rectangular well i.e. it is given by
\begin{eqnarray}
\left\{
\begin{array}{ll}
V(x,g) = 2g{\delta(x)} & {\mid x \mid < 1} \\[3mm]
\psi(\pm 1) = 0 &
\end{array}
\right.
\label{15}
\end{eqnarray}
where $\psi(x)$ is a wave function for the case. A role of a
perturbation parameter is played by g.  For $g=0$ we get a problem
of the energy spectrum in the infinite rectangular well i.e. an
asymptotic limit of any energy level $E_{n}(g)$ of the potential (\ref{14})
is just a corresponding level $E_{n}$ of the rectangular well. In the
limit $g \rightarrow +\propto$ we get instead two infinite rectangular wells 
with the half of that size of the well we started with for which their
energy spectra should coincide with the corresponding limit of
$E_{n}(g)$. This is not unexpected. What is interesting in this
example is the fact \cite{4}  that every even parity level of the
potential (\ref{14}) is a branch of some ramified function $E(g)$
considered as a function of complex {\it g} (odd parity energy levels
decouple of {\it g} and therefore coincide with the odd ones of the
infinite rectangular well). With respect to this property the
considered case reminds to some extent the previous one.
However there is one considerable difference between them. This
is that the existence of all the even energy levels $E_{n}(g)$ does
not depend on {\it g} i.e. each level exists for any {\it g}, including $g=0$.
Therefore, we cannot expect on a {\it g}-Riemann surface ${\bf R}_{g}$ 
of $E(g)$ the real branch points to appear having meanings of
pseudothresholds for the levels $E_{n}(g)$. On the other hand there
are branch points on ${\bf R}_{g}$ (as they have to be because $E(g)$ is a
ramified function) which are complex but which physical meaning
seems to be puzzled.  Nevertheless there is also a property of
the considered case which is a copy of the corresponding one for
the finite rectangular well. Namely, perturbative series for
even $E_{n}(g)$ are all convergent. All these follow from the
quantization condition for even energy levels
\begin{eqnarray}
\left\{
\begin{array}{l}
k \cot{k} = - g \\[5mm]
k = \sqrt{E}
\end{array}
\right.
\label{16}
\end{eqnarray}

The condition (\ref{15}) defines $g$ as a meromorphic function of {\it k} with
simple poles on the {\it k}-plane at $k=r{\pi}$, $r={\pm}1,{\pm}2$,.... 
Therefore an inversion of the relation (\ref{15}) will 
be easy if we know positions
of zeros of {\it g'(k)} in the {\it k}-plane. These positions are determined
by (\ref{15}) and by the following condition:
\begin{eqnarray}
g^2 + g + k^2 = 0
\label{17}
\end{eqnarray}

Both the last equations are equivalent to
\begin{eqnarray}
\left\{
\begin{array}{l}
\sin{2k} = 2k \\[5mm]
g = -k \cot{k}
\end{array}
\right.
\label{18}
\end{eqnarray}

The first of equations (\ref {17}) has a solution for $k=0$ and an
infinite number of complex solutions in the {\it k}-plane (i.e. Im $k \neq 0$ 
for all these solutions) with the property that if $k_{l}$,
$l=1,2$,..., is a solution to it then -$k_{l}$, $k_{l}^{*}$ and 
-$k_{l}^{*}$ are also.
It is easy to show that for large {\it l} Im$k_{l}$ increases as $\ln{l}$ and
Re$k_{l}$ as $l\pi$ \cite{4}. A corresponding pattern of lines 
Re{\it g}=const taking
into account the distribution of singular points described above
is shown in Fig. 9. The pattern allows us for an easy
reproducing of a {\it g}-Riemann surface ${\bf R}_{g}$ for an inverse function
$k(g)$.  Namely, we can arrange cuts on ${\bf R}_{g}$ in such a way to map a
$l^{th}$ dashed area of Fig. 9 into a sheet of ${\bf R}_{g}$ corresponding to
$E_{l}^{th}$ even energy level of the potential considered. This is
shown in Fig. 10.

\begin{tabular}{c}
\psfig{figure=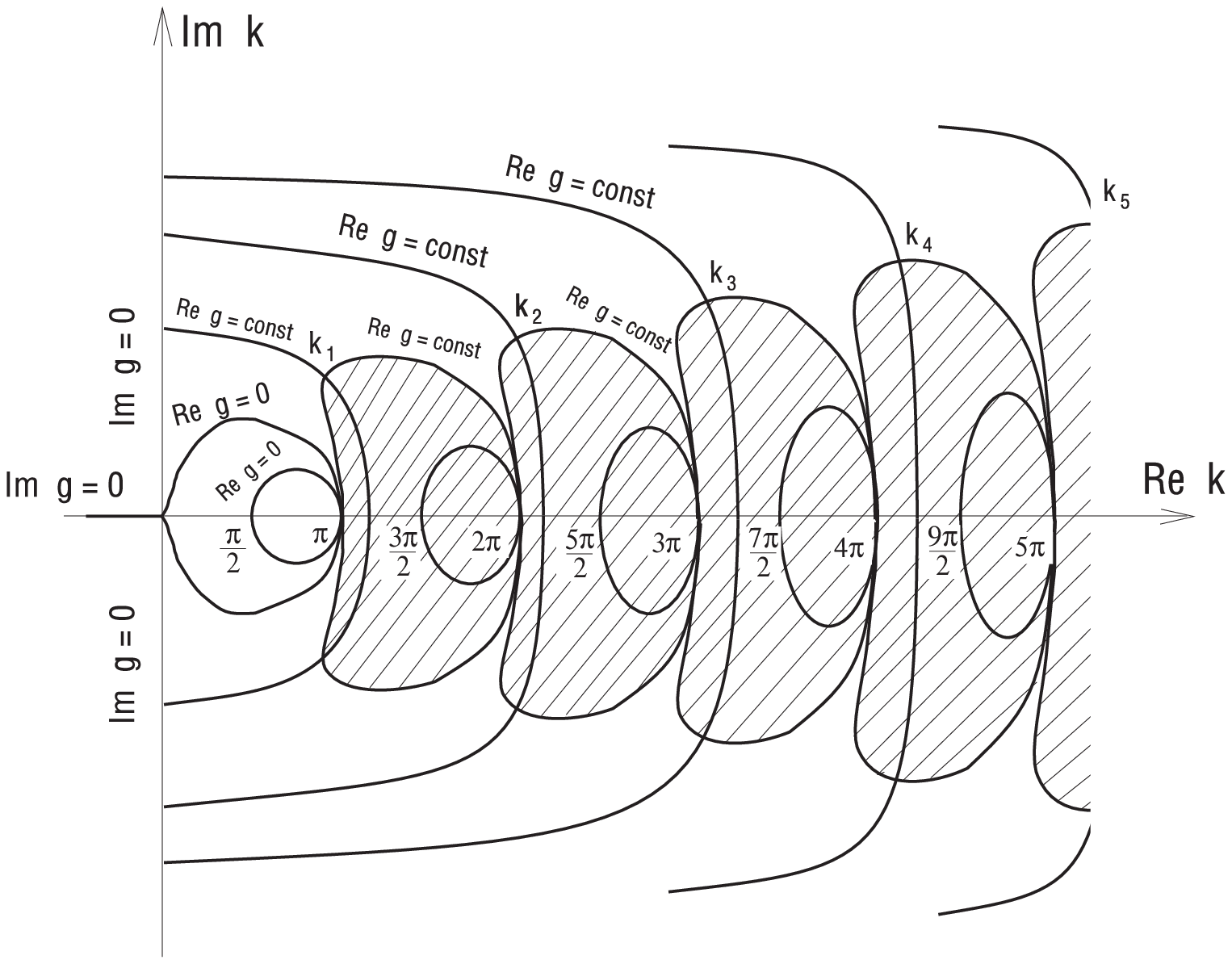,width=12cm} \\
\parbox{12cm}{Fig.9 The pattern of lines $Re g(k)=const$ and $Im g(g)=
const$ for $g(k) = -k \cot k $ }
\end{tabular}

\vspace{5mm}

Because $g(k)$ is a symmetric function of $k$
i.e. $g(k)=g(-k)$ with the singular point at $k=0$ (i.e. $g'(0)=0$)
its inverse function $k(g)$ is defined on the Riemann surface ${\bf R}_{g}$
which consists of two twin systems of sheets joined by one cut
only beginning at $g=-1$ and running to $-\propto$. These two systems of
sheets are maps of the right and the left {\it k}-halfplanes of Fig.9, 
respectively.  The two sheets opening the systems and joined
by the cut described above are the ones on which the ground
state energy level is defined taking the same values in the
corresponding points of the sheets. This is because $E(g)=k^{2}(g)$.
Therefore, for $E(g)$ the point $g=-1$ on the first sheet considered
is not a branch one and we can consider only one of the two
systems of sheets corresponding for example to Re$k \geq 0$ what is
assumed from now on.

In accordance with the results of
Ushveridze \cite{4} all singular points of $E(g)$ lie on the first
sheet corresponding to the ground state energy level of the
potential (\ref{15}). The sheet is an image of an area which completes
the dashed ones of Fig. 9 to the full right {\it k}-halfplane and is
sketched in Fig. 10.
\begin{center}
\unitlength=0.80mm
\special{em:linewidth 0.4pt}
\linethickness{0.4pt}
\begin{picture}(121.00,122.00)
\put(75.00,30.00){\line(0,1){90.00}}
\put(75.00,122.00){\makebox(0,0)[ct]{$\wedge$}}
\put(121.00,75.00){\makebox(0,0)[cc]{$>$}}
\put(84.00,122.00){\makebox(0,0)[cc]{Im g}}
\put(117.00,82.00){\makebox(0,0)[cc]{Re g}}
\put(51.00,97.00){\rule{1.00\unitlength}{23.00\unitlength}}
\put(51.00,31.00){\rule{1.00\unitlength}{22.00\unitlength}}
\put(58.00,97.00){\makebox(0,0)[cc]{$g(k_{1})$}}
\put(58.00,53.00){\makebox(0,0)[cc]{$\bar{g}_{1}$}}
\put(5.00,75.00){\line(1,0){115.00}}
\put(38.00,31.00){\rule{1.00\unitlength}{19.00\unitlength}}
\put(38.00,100.00){\rule{1.00\unitlength}{20.00\unitlength}}
\put(14.00,105.00){\rule{1.00\unitlength}{15.00\unitlength}}
\put(14.00,31.00){\rule{1.00\unitlength}{14.00\unitlength}}
\put(5.00,74.00){\rule{60.00\unitlength}{2.00\unitlength}}
\put(65.00,68.00){\makebox(0,0)[cc]{$g_{0}=-1$}}
\put(43.00,50.00){\makebox(0,0)[cc]{$\bar{g_2}$}}
\put(19.00,45.00){\makebox(0,0)[cc]{$\bar{g_l}$}}
\put(19.00,105.00){\makebox(0,0)[cc]{$g_l$}}
\put(43.00,100.00){\makebox(0,0)[cc]{$g_2$}}
\end{picture}
\\
Fig.10 The cut structure of the g-Riemann surface for k(g) determining
the surface
\end{center}

As it has been already explained earlier
on the $l^{th}$ sheet where the $l^{th}$ energy level $E_{l}(g)$ is defined on
there is a unique pair of complex conjugated branch points at 
$g_{l} (=g(k_{l}))$ and $\bar{g}_{l}$ by which the level contacts with the 
ground state one (see Fig. 11). The level is holomorphic on its sheet
at $g=0$ and approaches $[(l+1) \pi]^2$ or $(l \pi)^2$ for 
$g \rightarrow \propto$ 
in the halfplanes Re$g >$ Re$g_{l}$ or Re$g <$ Re$g_{l}$ respectively.

Because of the
holomorphicity of $E_{l}(g)$ at $g=0$ its perturbative series with
respect to $g$ converges inside the circle $\mid g \mid = \mid g_{l} \mid$.  

For large $g$ however we have to consider together with $E_{l}(g)$ also its
neighbour $E_{l+1}(g)$ for which its limit for $g$ in the halfplane
Re$g <$ Re$g_{l}$ is the same as for $E_{l}(g)$ in the halfplane 
Re$g >$ Re$g_{l}$. A common sheet for both the levels is shown in Fig. 12. 
It is obvious that we can apply here the Cauchy integral technique to
show the convergence of the asymptotic series for $E_{l}(g)$ and for
$E_{l+1}(g)$ for large g with 
$\mid g \mid > \mid g_{l+1} \mid$, $l=1,2$,.., .  

\begin{eqnarray*}
\unitlength=0.59mm
\special{em:linewidth 0.4pt}
\linethickness{0.4pt}
\begin{picture}(121.00,122.00)
\put(75.00,30.00){\line(0,1){90.00}}
\put(75.00,122.00){\makebox(0,0)[ct]{$\wedge$}}
\put(30.00,75.00){\line(1,0){90.00}}
\put(121.00,75.00){\makebox(0,0)[cc]{$>$}}
\put(84.00,122.00){\makebox(0,0)[cc]{Im g}}
\put(117.00,82.00){\makebox(0,0)[cc]{Re g}}
\put(51.00,97.00){\rule{1.00\unitlength}{23.00\unitlength}}
\put(51.00,31.00){\rule{1.00\unitlength}{22.00\unitlength}}
\put(94.00,80.00){\makebox(0,0)[cc]{$E_{l}(g)$}}
\put(50.00,80.00){\makebox(0,0)[cc]{$E_{l}(g)$}}
\put(58.00,97.00){\makebox(0,0)[cc]{$g(k_{l})$}}
\put(58.00,53.00){\makebox(0,0)[cc]{$\bar{g}(k_{l})$}}
\end{picture}
&
\unitlength=0.58mm
\special{em:linewidth 0.4pt}
\linethickness{0.4pt}
\begin{picture}(122.00,122.00)
\put(75.00,30.00){\line(0,1){90.00}}
\put(75.00,122.00){\makebox(0,0)[ct]{$\wedge$}}
\put(122.00,75.00){\makebox(0,0)[cc]{$>$}}
\put(84.00,122.00){\makebox(0,0)[cc]{Im g}}
\put(117.00,82.00){\makebox(0,0)[cc]{Re g}}
\put(58.00,97.00){\makebox(0,0)[cc]{$g(k_{l})$}}
\put(58.00,53.00){\makebox(0,0)[cc]{$\bar{g}(k_{l})$}}
\put(5.00,75.00){\line(1,0){115.00}}
\put(50.00,55.00){\rule{1.00\unitlength}{40.00\unitlength}}
\put(30.00,40.00){\rule{1.00\unitlength}{70.00\unitlength}}
\put(37.00,111.00){\makebox(0,0)[cc]{$g_{l+1}$}}
\put(38.00,39.00){\makebox(0,0)[cc]{$\bar{g_{l+1}}$}}
\put(8.00,81.00){\makebox(0,0)[cc]{$E_{l+1}(g)$}}
\put(69.00,81.00){\makebox(0,0)[cc]{$E_{l}(g)$}}
\end{picture}
\\
\parbox{7.2cm}{Fig.11 The $g$-Riemann surface for $l^{th}$
energy level of the infinite
rectangular well with the Dirac delta barrier}
&
\parbox{7.2cm}{Fig.12 The cut structure of the sheet
on which the perturbative series
for $E_{l}(g)$ is studied}
\end{eqnarray*}

The corresponding
statement for the ground state energy level perturbation series
is also true for $\mid g \mid > \mid g_{1} \mid$.

\section{Conclusions} 

The model of finite rectangular well which we
have considered provided us with several properties differing it
from the previous ones. First it was a number of energy levels
varying with a perturbation parameter thus changing their
quality of being real to becoming only potential ones.  

Secondly it was the existence of the pseudoenergy levels in the model the
main role of which was just to allow the level crossing
phenomenon to appear.

It was also an easiness of predictions of
the pseudothreshold distribution as a part of the level crossing
loci. The latter property followed however from the simplicity
of the quantization conditions for the case as illustrated by
Fig. 1. In fact the remaining three branch points, the two for
the ground state level at $\lambda = \pm i$ and the logarithmic one for all
the levels at $\lambda = 0$ could be identified only by detailed
calculations.  

Finally it was the convergence of perturbation
series for large $\lambda$ and for all the levels.  This property could
not be deduced prior to the detailed knowledge of the $\lambda$-Riemann
surface topology just because for imaginary $\lambda$ (no matter small or
large) the rectangular well became the rectangular barrier
suggesting rather a possibility for the perturbation series to
be divergent in these directions because of a repelling
character of the barrier. However the repulsion of the
rectangular barrier appears to be not enough strong to destroy the
convergence of the perturbation series.  

The property of the perturbation series to be convergent was shared also 
by the corresponding series constructed for the levels in the infinite
rectangular well perturbed by the Dirac delta barrier.

Generally the analytical properties of energy levels considered
above show that they depend strongly on the model used. Crossing
of levels considered as the main cause for the presence of
branch points in the dependence of levels on a chosen
perturbation parameter seems to be as obvious as puzzled in most
of the models. In particular these are the positions of branch
points in investigated models which does not seem to be covered
by some universal rules. This latter statement is true also in
the case of the rectangular well energy levels investigated in
this paper and the name of pseudothresholds we gave to the
underlying branch points cannot be considered more seriously
than only a convenient convention. It seems rather that such a large
variety of different parameters which were used as perturbation
ones in models considered so far does not allow us to formulate some
common rules covering the distributions of branch points and
giving them in this way some common sense. Only the most general
hints such as for example the non existing of energy level
degeneracy in the 1-dim quantum mechanics can help us to make
some predictions.  In most cases however only a detailed
knowledge of considered potentials and corresponding
quantization conditions the former generate can allow us to
formulate some reasonable predictions prior to detailed
calculations.

\end{document}